\documentclass{book}
\usepackage{latexsym}
\usepackage{amsmath}
\usepackage{bm}
\makeatletter
\def\my@tag@font{\normalsize}
\def\maketag@@@#1{\hbox{\m@th\normalfont\my@tag@font#1}}
\let\amsmath@eqref\eqref
\renewcommand\eqref[1]{{\let\my@tag@font\relax\amsmath@eqref{#1}}}
\makeatother
\usepackage{amssymb}
\usepackage{color}
\usepackage{colortbl}
\definecolor{mygray}{gray}{.9}
\definecolor{mypink}{rgb}{.99,.91,.95}
\definecolor{mycyan}{cmyk}{.3,0,0,0}
\usepackage{mathrsfs} %special symbol package

\usepackage[paperwidth=21cm,paperheight=29.7cm]{geometry}
\usepackage[labelsep=period, labelfont=bf, aboveskip=10pt, belowskip=2pt, font=small, font=it]{caption}

\usepackage[english]{babel}
\emergencystretch=\maxdimen
\usepackage{indentfirst} %Indent the first line of a paragraph
\usepackage{ifthen}
\usepackage{amsmath}
\usepackage{fancyhdr}
\usepackage{amsthm}
\usepackage[normalem]{ulem}

\usepackage[latin1]{inputenc}
\usepackage{amsfonts}
\usepackage{amssymb}
\usepackage{multicol}
\usepackage{makeidx}
\usepackage{graphicx}
\usepackage{ulem} % is used for different types of under lining
\usepackage{subcaption}
\usepackage[numbers,sort&compress]{natbib}
\usepackage{setspace}
\fancypagestyle{plain}{\fancyhf{}}

\usepackage{titlesec}
\titlelabel{\thetitle.~~}
\titlespacing*{\subsubsection}{0pt}{12pt}{0pt}
\titleformat{\subsection}{\bfseries\slshape}{\thesubsection.}{0.5em}{}
\titleformat{\subsubsection}{\bfseries\slshape}{\thesubsubsection.}{0.5em}{}

\usepackage{times}
\usepackage{calc}
\usepackage{xcolor}
\usepackage{amsmath}
\usepackage{amssymb}
\usepackage{array}
\usepackage{booktabs}
\usepackage{caption}
\usepackage{longtable}
\usepackage{tabularx}
\usepackage{fancyvrb}
\usepackage{graphicx}
\usepackage{epstopdf}
\usepackage{ifthen}
\usepackage{paralist}
\usepackage{titletoc}
\usepackage{ulem}
\usepackage{float}
\usepackage{latexsym}

\usepackage{dsfont}
\usepackage{makeidx}
\usepackage{lipsum}
\usepackage{imakeidx}
\usepackage{multirow}

\usepackage[hang]{footmisc} %make footnotes indent a fixed width
\makeatletter
\def\thm@space@setup{\thm@preskip=0pt
\thm@postskip=0pt}
\makeatother
\newtheoremstyle{remark}
{} %Aboveskip
{} %Below skip
{\mdseries} %Body font e.g.\mdseries,\bfseries,\scshape,\itshape
{} %Indent
{\bfseries} %Head font e.g.\bfseries,\scshape,\itshape
{.} %Punctuation afer theorem header
{} %Space after theorem header
{} %Heading
\theoremstyle{remark}


\begin{document}
\renewcommand{\qedsymbol}{}
\DeclareFixedFont{\Head}{OT1}{phv}{bx}{n}{18pt}
\DeclareFixedFont{\SEC}{OT1}{phv}{bx}{n}{12pt}
\DeclareFixedFont{\Journal}{OT1}{phv}{bx}{n}{11pt}
\DeclareFixedFont{\Name}{OT1}{ptm}{bx}{n}{12pt}%
\DeclareFixedFont{\Address}{OT1}{ptm}{m}{n}{9pt}
\DeclareFixedFont{\Emailaddress}{OT1}{ptm}{bx}{n}{12pt}
\DeclareFixedFont{\Emailaddresscontent}{OT1}{ptm}{m}{n}{9pt}%
\DeclareFixedFont{\Citation}{OT1}{ptm}{bx}{n}{12pt}
\DeclareFixedFont{\Citationcontent}{OT1}{ptm}{m}{n}{9pt}
\DeclareFixedFont{\Date}{OT1}{ptm}{m}{n}{9pt}
\DeclareFixedFont{\Abstract}{OT1}{ptm}{bx}{n}{12pt}
\DeclareFixedFont{\Abstractcontent}{OT1}{ptm}{m}{n}{10pt}
\DeclareFixedFont{\FirstLevel}{OT1}{ptm}{bx}{n}{14pt}
\DeclareFixedFont{\SecondLevel}{OT1}{ptm}{bx}{it}{10pt}
\DeclareFixedFont{\ThirdLevel}{OT1}{ptm}{bx}{it}{10pt}%
\DeclareFixedFont{\Text}{OT1}{ptm}{m}{n}{10pt}
\DeclareFixedFont{\References}{OT1}{ptm}{bx}{n}{14pt}%
\DeclareFixedFont{\ZJL}{T1}{pxtt}{bx}{n}{12pt}
\chapter*{}
\setcounter{page}{28}%
\vspace{-4.65cm}
%\makebox[48.6em][l]{
\noindent
\begin{tabular}[H]{lr}\toprule [1pt] \vspace{-0.33cm}\\
\hspace{-2mm}\Journal{American Journal of Physics and Applications} & \hspace{30.5mm}\multirow{5}*{\includegraphics[height=22mm,width=58mm]{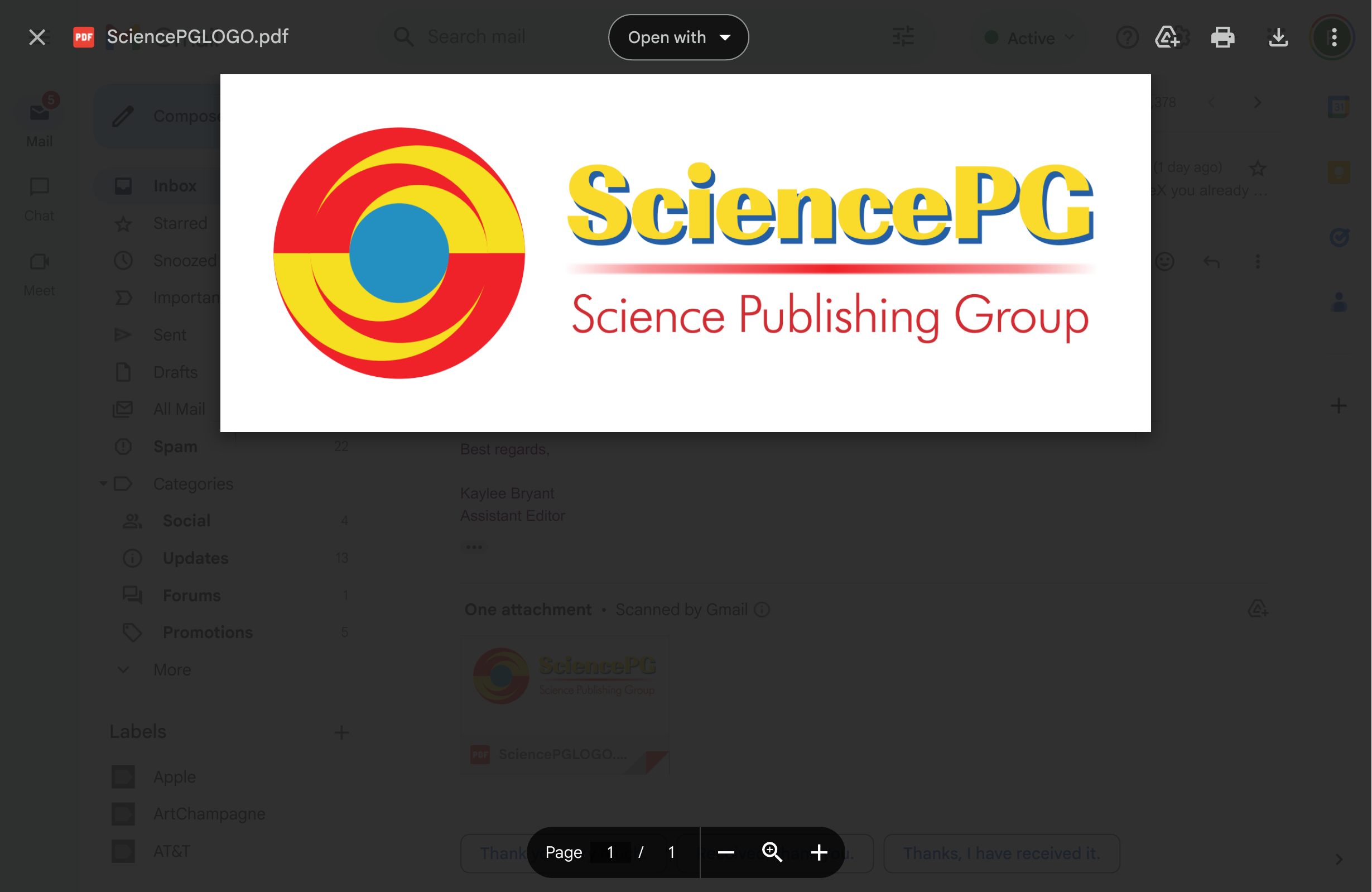}}\\
\hspace{-2mm}\Address{2025; 13(2): 28-30} &\\
\hspace{-2mm}\Address{http://www.sciencepublishinggroup.com/j/ajpa} &\\
\hspace{-2mm}\Address{doi: 10.11648/j.ajpa.20251302.12} &\\
\hspace{-2mm}\Address{ISSN: 2330-4286 (Print); ISSN: 2330-4308 (Online)} \\\bottomrule [2.5pt]
\end{tabular}
%}
\parskip=12pt
\begin{flushleft}
\noindent \Head{A Physics Model for Origin of Life}
\end{flushleft}
\parskip=8pt
\noindent \Name{Paul Howard Frampton}
\parskip=8pt

\noindent \Address{Department of Mathematics and Physics ``Ennio De Giorgi'', University of Salento and INFN-Lecce, Lecce, Italy}
\vspace{5pt}

\parskip=4pt
\noindent \Emailaddress{Email address:}\\
\noindent \Emailaddresscontent{paul.h.frampton@gmail.com}
%\noindent \Emailaddresscontent{lanyuyanliang@163.com}

\parskip=8pt
\noindent \Citation{To cite this article:}
\begin{flushleft}
\vspace{-0.56cm}
\noindent \Citationcontent{Paul Howard Frampton. (2025). A Physics Model for Origin of Life. {\it American Journal of Physics and Applications, 13}(2), 28-30. https://doi.org/10.11648/j.ajpa.20251302.12}
\vspace{-0.3cm}
\end{flushleft}
\parskip=10pt

\noindent \Date{{\bf Received:} 17 February 2025; {\bf Accepted:} 5 March 2025; {\bf Published:} 10 April 2025}

\parskip=5pt
\noindent \rule{\textwidth}{1pt}

\parskip=6pt
\noindent \Abstract{Abstract: }\Abstractcontent{\baselineskip=12pt In this article, we attempt to convince the reader that the origin of life was such an exceptionally unlikely event that it probably has never occurred elsewhere. This involves an explicit calculation using the laws of physics which, while speculative, may encapsulate the essential science without knowledge of biological details. Making only physics, and no biology, assumptions about the origin of the first single celled organism (SCO) on Earth, we adopt methods of quantum tunnelling to make an estimate of the probability ${\cal P}_{SCO}$ for the origin of life. We argue that before the time $t_{SCO}$ laws of physics must suffice and assume a first-order phase transition which nucleates at the first SCO production. In the classical limit where Planck's constant vanishes $h\rightarrow0$, ${\cal P}_{SCO}$ also vanishes and remains extremely small for the correct value of $h$. Thus quantum mechanics plays a central role in permitting life to form. We compare the resultant probability with the expected number of exoplanets in the Milky Way ($\sim10^{12}$) and the Visible Universe ($\sim10^{24}$) and conclude that the probability of extraterrestrial life in the Visible Universe is infinitesimal. This result suggests that the visible universe is a lonely place for humankind because extraterrestrial life will never be encountered.}

\parskip=8pt
\hangafter=1
\setlength {\hangindent}{6.2em}
\noindent \Abstract{Keywords: }\Abstractcontent{Vacuum Tunnelling, Planck's Constant, First-order Phase Transition, Origin of Life}
\parskip=3pt

\noindent \rule{\textwidth}{1pt}
\parskip=0pt
%double column display
\columnseprule=0pt
\setlength{\columnsep}{1.5em}
\vspace{-12pt}
%\vspace{-7pt}
\begin{multicols}{2}
\section{Introduction}
\vspace{-6pt}
%\vspace{-11pt}
As we begin the centennial year for the discovery of quantum mechanics, we cannot resist the temptation to explore the application of quantum mechanics outside of the inanimate objects studied in physics. This leads us into a research field of quantum biology. Previous attempts to apply quantum mechanics to biology by Edelman \cite{Edelman} and Penrose \cite{Penrose} have focused on the human brain and consciousness. Here our aim seems less ambitious, but is nevertheless extremely challenging. It is to discuss the origin of life on the Earth. One definition of life is that it is capable of reproducing itself whereas inanimate objects cannot.

We aim to estimate the probability ${\cal P} _{SCO}$ of formation at time $t_{SCO}$ of the first Single Celled Organism (SCO) on the Earth. In the absence of biology prior to $t_{SCO}$, we are permitted to use physics without biology. Our assumption will be that the physics will involve a First Order Phase Transition (FOPT) triggered by nucleation at the time and place of the first SCO. A FOTP is one of the most violent transitions in physics. In quantum field theory, the FOPT transition from a false unstable vacuum to a real stable one provides a suitable formalism and we identify the situation without life with the false vacuum and that with life as the stable one. We shall defer any discussion of the probability ${\cal P}_{fission}$ for the SCO to fission into two of itself.

We shall characterise this by the exponent $n$ in ${\cal P}_{SCO} \equiv 10^{-n}$. Given that there are of order $\sim 10^{12}$ exoplanets in the Milky Way, and $\sim 10^{24}$ in the visible universe, we expect that if $n>24$ there is no extraterrestrial life. On the other hand, if $12 < n < 24$ or $0 < n < 12$ respectively, we may expect extraterrestrial life in the Universe or even in the local Milky Way galaxy. We have adopted the conventional estimates for the numbers of stars and exoplanets in a typical galaxy and that there are approximately a trillion galaxies in the visible universe. These order of magnitude estimates suffice to reach our principal conclusion.

The time and place are not fully established. The time $t_{SCO}$ is widely believed to be between 4.0 Gy and 4.2 Gy ago \cite{Bell, Dodd}, just a few hundred million years after the time $t_{\oplus}$ of the formation of the Earth at $t_{\oplus} \simeq 4.54Gy$ ago \cite{Dalrymple, Manhesa, Braterman}. The place is usually believed to be below the surface of an ocean, perhaps in or near to a thermal vent emanating from graphitic rock.

Fortunately we shall not need to know when and where the first SCO appeared on the Earth to make our calculation, nor shall we use any knowledge about biology.

\section{The Calculation}
Our discussion of ${\cal P}_{SCO}$ will be by analogy with the first-order phase transition (FOPT) of vacuum decay in quantum field theory \cite{Frampton, Coleman, Callan}, with no knowledge of biology.

One scenario for the FOPT is the following topological transition, We assume inanimate long organic molecules (LOMs) including amino acids, nucleic acids (both DNA and RNA) are already assembled into a planar rectangular sheet of cell membrane. This sheet can be rolled into a cylinder, then the cylinder closed upon itself into a genus $g=1$ torus, The genus is classically a topological invariant. To arrive at an SCO requires quantum tunnelling to a genus $g=0$ surface with the same topology as the surface of a sphere.\footnote{This specific topological FOPT was suggested by T. Kephart.} We should emphasise that we believe the subsequent FOPT calculation is of greater generality than this topological example. There may well exist a better FOPT description than the one briefly described in the present paragraph.

We take the $g=1$ state to be an unstable vacuum which can quantum tunnel to decay by spherical bubble nucleation into the stable $g=0$ vacuum. The energy ${\cal E}$ for the bubble with radius $R$ is
\begin{equation}
{\cal E} = - \frac{1}{2} \pi^2 R^4 \epsilon + 2 \pi^2 R^3 \sigma
\label{energy}
\end{equation}

\noindent where
\begin{equation}
\epsilon = \epsilon_{g=1} - \epsilon_{g=0}
\label{epsilon}
\end{equation}

\noindent is the volume energy density of the bubble in a thin-wall approximation and $\sigma$ is the surface energy density.

Stationarising ${\cal E}$ in Eq. (\ref{energy}) by setting $d{\cal E}/dR=0$ gives a critical radius $R_m=3\sigma/\epsilon$ and a minimum energy
\begin{equation}
{\cal E}_m = 27 \pi^2 \left(\frac{\sigma^4}{\epsilon^3} \right)
\label{Am}
\end{equation}

\noindent so that, temporarily ignoring a subdominant pre-factor, the probability of quantum tunnelling is
\begin{equation}
{\cal P}_{tunnelling} = \exp(-{\cal E}_m) = 10^{-n}
\label{calP}
\end{equation}

\noindent which implies that
\begin{equation}
n = \left(\frac{{\cal E}_m}{ln 10} \right)
\label{n}
\end{equation}

Actually, the energy in Eq. (\ref{energy}) from reference \cite{Frampton} is Lorentz invariant and special relativistic. For the origin of life, we need only the non-relativistic simplification
\begin{equation}
{\cal E} = - \frac{4 \pi}{3} R^3 \epsilon + 4 \pi R^2 \sigma
\label{NRenergy}
\end{equation}
\vspace{0.3cm}
\noindent which is stationary for $R_m=2 \sigma / \epsilon$ with the value
\begin{equation}
{\cal E}_m = \frac{16 \pi}{3} \left(\frac{\sigma^3}{\epsilon^2} \right)
\label{NREm}
\end{equation}

\noindent so that
\begin{equation}
n \equiv \frac{{\cal E}_m}{\ln 10} = \frac{16 \pi}{3 \ln 10}\left(\frac{\sigma^3}{\epsilon^2} \right) = 7.277 \left(\frac{\sigma^3}{\epsilon^2} \right)
\label{NRn}
\end{equation}

In order to evaluate $n$ in Eq. (\ref{NRn}) we must first restore the units from the use of $h=c=1$ natural units. This means that the full expression becomes
\begin{equation}
n = 7.277 \left(\frac{\sigma^3}{h \epsilon^2} \right) \sim 9.30 \times 10^{31}
\label{nfinal}
\end{equation}

\noindent where, in kilogram-metre-second units, we used $h= 6.262\times 10^{-34}$; $\sigma=2\times 10^{-3}$; and (estimated) $\epsilon=10^{-3}$.

With this result, our estimate for the probability of the creation of a Single Celled Organism ${\cal P}_{SCO}$ becomes
\begin{equation}
{\cal P}_{SCO} \sim 10^{-9.3\times10^{31}}
\label{finalPsco}
\end{equation}

\noindent which is so tiny that our conclusion about extraterrestrials is rendered black and white. The subdominant pre-factor, hitherto ignored, is irrelevant and the question of the Milky Way versus the Visible Universe also become moot, because ${\cal P}_{PCO}$ in Eq. (\ref{finalPsco}) when multiplied by $10^{12}$ or $10^{24}$ remains infinitesimal as does the probability of extraterrestrial life.

Because of the small size of a SCO, the time taken by the FOPT with probability estimated in Eq. (\ref{finalPsco}) is expected to be measured in seconds rather than years.

We have assumed that for times satisfying $t < t_{SCO}$, when there was no life, there was no biology and the laws of physics must suffice. Whether this assumption continues to hold for all later times $t > t_{SCO}$ is debatable. For example, at the present time there is the phenomenon of human consciousness which some, including the present author, suspect will remain forever mysterious but other scientists expect human consciousness will be fully explained by neuroscience using the laws of physics.

\section{Non-Zero Temperature}
In quantum mechanical examples the suppression created by quantum tunnelling can, in some cases, be alleviated by a high ambient temperature, so we need to discuss whether this is likely to change our estimate of ${\cal P}_{SCO}$ in Eq. (\ref{finalPsco}).

If, for example, the first SCO was created in a thermal vent under the ocean it is safe to assume the extant temperature was below the boiling point of water at 373K.

Using Boltzmann's constant k the conversion from atomic energy units is 1eV = 11,600K so that the binding energy within an H atom is 13.6 eV = 158,000 K which is the approximate scale of background temperature needed to obviate quantum tunnelling, and is orders of magnitude higher than 373K. This suggests that nonzero temperature does not significantly change our estimate in Eq. (\ref{finalPsco}).

It is unknown whether the formation of the first SCO required a pressure greater than normal atmospheric pressure, for example by being deep under water. However, we expect pressure, like temperature, to provide only a small alteration in Eq. (\ref{finalPsco}).

\section{Discussion}
Ours appears as a reasonable physics model for the origin of life and addresses a question which is of importance in the history of the humankind and widely regarded as a question for biology, not physics. It was certainly an abrupt and extreme change of which the best physics example is a first-order phase transition. In everyday life, FOPTs from ice to water, and from water to steam, are familiar.

The appearance of Planck's constant in the denominator of Eq. (\ref{nfinal}) shows that, in the classical limit $h\rightarrow0$, ${\cal P}_{SCO}\rightarrow0$ as expected. The placement of $h$, and its smallness, is the reason in the above calculation why the creation of life is so improbable.

Why should we be interested in ${\cal P}_{SCO}$? We may have been entertained by the movie "ET, the Extra Terrestrial" and wondered whether anything like that could ever really happen. Our conclusion is that it was purely fictional.

More than anything, our estimate of ${\cal P}_{SCO}$ in Eq. (\ref{finalPsco}) suggests that, for humankind, the universe is a lonely place.

\section*{Conflicts of Interest}
The author declares no conflict of interest.

%\bibliographystyle{ieeetr}
%\bibliography{WGE}
\parskip=5pt
\noindent \rule[0pt]{25em}{1pt}

\parskip=10pt
\noindent {\References{References}}
\vspace{-6pt}
\begin{list}{[\arabic{enumi}]}{\usecounter{enumi} %%
\leftmargin=2em%%%%%%
\labelsep=0em%%%%%
\renewcommand{\makelabel}[1]{#1 \hfil}}%%
\bibitem{Edelman} G. Edelman, {\em Neural Darwinism}. Basic Books, New York. (1987).
\bibitem{Penrose} R. Penrose, {\em The Emperor's New Mind}. Oxford University Press (1989).
\bibitem{Bell} E. A. Bell, P. Boehnke, T, M, Harrison, {\em et al.}, {\em Potentially Biogenic Carbon Preserved in a 4.4 Billion-Year-Old Zircon.} Proceedings of the NAS, {\em 112}, 14518 (2015). https://doi.org/10.1073/pnas.1517557112
\bibitem{Dodd} M. S. Dodd, D. Papinau, T. Grenne, {\em at al.} {\em Evidence for Early Life in Earth's Oldest Hydrothermal Vent Precipitates.} Nature {\em 543}, 66 (2017). https://doi.org/10.1038/nature21377
\bibitem{Dalrymple} G. B. Dalrymple, {\em The Age of the Earth on the Twentieth Century: a Problem (Mostly) Solved.} Special Publications, Geological Society of London, {\em 190}, 205 (2001). https://doi.org/10.1144/GSL.SP.2001.190.01.14
\bibitem{Manhesa} G. Manhesa, C. J. Allegre, B. Duprea and B. Hamelin, Lead Isotope Study of Basic-Ultrabasic Layered Complexes: Speculations about the Age of the Earth and Primitive Mantle Characteristics. Earth and Planetary Science Letters, {\em 47,} 370 (1980). https://doi.org/10.1016/0012-821X(80)90024-2
\bibitem{Braterman} P. S. Braterman, {\em How Science Figured Out the Age of the Earth.} Scientific American (October 2013).
\bibitem{Frampton} P. H. Frampton, {\em Vacuum Stability and Higgs Scalar Mass.} Phys. Rev. Lett. {\em 37,} 1378 (1976). https://doi.org/10.1103/PhysRevLett.37.1378
\bibitem{Frampton} P. H. Frampton, {\em Vacuum Instability in Quantum Field Theory.} Phys. Rev. {\em D15}, 2922 (1977). https://doi.org/10.1103/PhysRevD.15.2922
\bibitem{Coleman} S. Coleman, {\em Tha Fate of the False Vacuum. I. Semiclassical Theory.} Phys. Rev. {\em D15}, 2929-34 (1977). https://doi.org/10.1103/PhysRevD15.2829
\bibitem{Callan} C. G. Callan and S. R. Coleman, {\em Tha Fate of the False Vacuum. II. First Quantum Corrections}, Phys. Rev. {\em D16}, 1762-68 (1977). https://doi.org/10.1103/PhysRevD16.1762\\
\end{list}

\end{multicols}
\end{document}